\documentclass[1p]{elsarticle}

\usepackage{amssymb,amsthm}
\usepackage{graphicx}

\begin{document}

\begin{frontmatter}

\title{Characterization in bi-parameter space of a non-ideal oscillator}

\author{S. L. T. de Souza$^{1,2,\ast}$, A. M. Batista$^{3}$,
M. S. Baptista$^4$, I. L. Caldas $^2$ and J. M. Balthazar$^{5}$}
\address{$^1$Departamento de F\'isica e Matem\'atica, Universidade Federal de 
S\~ao Jo\~ao del Rei, Caixa Postal 131, 36420-000, Ouro Branco, MG, Brazil}
\address{$^2$Instituto de F\'isica, Universidade de S\~ao Paulo, 05315-970, 
S\~ao Paulo, SP, Brazil}
\address{$^3$Departamento de Matem\'atica e Estat\'istica, Universidade
Estadual de Ponta Grossa, 84030-900, Ponta Grossa, PR, Brazil}
\address{$^4$Institute for Complex Systems and Mathematical Biology,
SUPA, University of Aberdeen, AB24 3UE Aberdeen, United Kingdom}
\address{$^5$ ITA (Aeronautics Technological Institute), Mechanical-Aeronautics
Division, 12228-900, S\~ao Jos\'e dos Campos, SP, Brazil}

\cortext[cor]{Corresponding author: thomaz@ufsj.edu.br}

\date{today}

\begin{abstract}
We investigate the dynamical behavior of a non-ideal Duffing oscillator, 
a system composed of a mass-spring-pendulum driven by a DC motor with limited
power supply. To identify new features on Duffing oscillator parameter space
due to the limited power supply, we provide an extensive numerical
characterization in the bi-parameter space by using Lyapunov exponents. 
Following this procedure, we identify remarkable new periodic windows, the ones
known as Arnold tongues and also shrimp-shaped structures. Such windows appear 
in highly organized distribution with typical self-similar structures for the
shrimps, and, surprisingly, codimension-2 bifurcation as a point of 
accumulations for the tongues.
\end{abstract}

\begin{keyword}
Chaos \sep Arnold tongues \sep Shrimps \sep Coupled oscillators
\end{keyword}

\end{frontmatter}

\section{Introduction}

In recent years, there has been an increasing amount of work on nonlinear 
dynamics characterizing the possible structures in two-dimensional control 
parameter (bi-parameter) space \cite{gallas-1993}. Accordingly, 
periodic windows with important features, mainly shrimp-shaped structures 
\cite{gallas-1994} and Arnold tongues \cite{ecke1989,hengtai2013,pereira2014}, 
have been identified in several systems such as two-gene model  
\cite{desouza-2012}, impact oscillator \cite{desouza-2009,medeiros-2010}, 
dissipative model of relativistic particles \cite{denis}, 
tumor growth model \cite{stegemann-2014}, Chua's circuit 
\cite{baptista-2003,albuquerque-2008,hoff-2014}, prey-predator model 
\cite{rosa-2009}, and Red Grouse population model \cite{slipantschuk-2010}.

In the nonlinear dynamics context, oscillators with mechanical coupling have 
recently attracted a significant attention due to the complexity of the 
dynamics for high degree-of-freedom devices and possible applications to 
advanced technologies 
\cite{iliuk-2014,lenci-2011,chavez-2013,liu-2013,silva-2013}. 
Among the class of mechanical coupling oscillators an interesting example is 
the mass-spring-pendulum system 
\cite{sheheitli-2011,banerjee-1996}. 
Svoboda and collaborators studied a system of masses with a pendulum, where
the pendulum is attached to one mass of a chain of masses connected by
springs \cite{svoboda-1994}. They showed that auto parametric resonance can
arise. In Reference \cite{warminski-2005} was investigated the influence
of nonlinear spring on the auto parametric system. It was verified the existence
of regular and chaotic motions.

In this work, we investigate the parameter space organization of a non-ideal 
Duffing oscillator, namely, the mass-spring-pendulum system. Duffing oscillator
is a for\-ced oscillator with a nonlinear elasticity, and it is described
by a non linear differential equation of second-order that has been used in
a variety of physical processes. This oscillator is 
well known in engineering science, and it has been used to model the dynamics of
types of electrical and mechanical systems. Almong and collaborators 
experimentally studied signal amplification in a nanomechanical Duffing 
resonator via stochastic resonance \cite{almong2007}. The Duffing oscillator is
also a useful model to study the dynamics behavior of structural systems, such 
as columns, gyroscopes, and bridges \cite{suhardjo1992}. 

The non-ideal character of the studied oscillator is a consequence of the fact 
that the source of energy is given by a DC motor with limited power supply 
\cite{balthazar-2003,goncalves-2014}. Previous studies of this system have 
shown a rich dynamical behavior with several nonlinear phenomena, like 
quasi-periodic attractors, chaotic regimes, crises, coexistence of attractors, 
and  fractal basin boundaries \cite{desouza-2006,desouza-2007a,desouza-2008}. 
Here, our main purpose is to provide a global parameter analysis of the 
behavior of this oscillator with a mechanical coupling. The main features found
in the parameter space were the self-similar structures, and codimension-2 
bifurcation as a point of accumulations for the Arnold tongues. Comparing with 
results from parameter spaces of ideal oscillators, these Arnold tongue 
attributes are a consequence of the non-ideal character of this oscillator. 

This paper is organized as follows. In Section 2 we present the mathematical 
description of the non-ideal Duffing oscillator. In Section 3, we show our 
numerical analysis. The last section contains our main conclusions.

\section{Non-ideal Duffing oscillator}

\begin{figure}[hbt] 
\centering
\includegraphics[scale=0.5]{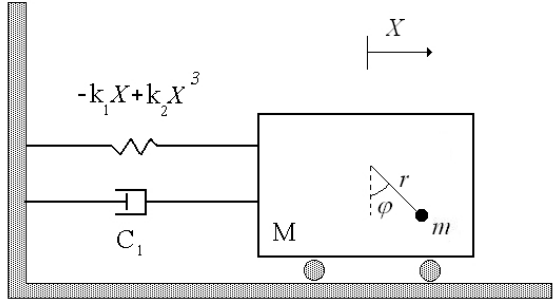}
\caption{\label{fig1} Schematic model of the non-ideal oscillator.}
\end{figure}

Several mechanical systems can be described by the Duffing equation. Tusset and 
Balthazar \cite{tusset2013} studied ideal and non-ideal Duffing oscillator 
with chaotic behavior. Th\-ey suppressed the chaotic oscillations through the 
application of two control signals. In this work, we consider a non-ideal 
system consisting of a mass, spring and pendulum. Figure \ref{fig1} shows a 
schematic model of the non-ideal oscillator \cite{desouza-2008}, that is 
composed of a cart (mass $M$), with a pendulum (mass $m$ and length $r$), 
connected to a fixed frame by a nonlinear spring and a dash-pot. We denote by 
$X$ the displacement of the cart and by $\varphi$ the angular displacement of 
the pendulum. 
  
\pagebreak

The equations of motion, obtained by using Lagrangian approach, for both the 
cart and the pendulum are given by:
\begin{eqnarray}\label{equation1}
(m+M)\frac{d^2 X}{d t^2} + c_1 \frac{d X}{d t} - k_1 X + k_2 X^3  &=&  
m r \left(\frac{ d \varphi}{d t}^2 \sin\varphi - \frac{d^2 \varphi}{d t^2} 
\cos\varphi \right), \\
m r^2 \frac{d^2\varphi}{d t^2} + c_2 \frac{d \varphi}{d t} 
+ m gr  \sin\varphi &=& 
E - m r\frac{d^2X}{d t^2} \cos\varphi, 
\end{eqnarray}

\noindent where $E$ is a constant source of energy. According to Equation 
(\ref{equation1}), for $k_1<0$, the Duffing oscillator can be interpreted as a 
forced oscillator with a spring whose restoring force is $F=k_1X-k_2X^3$. 
Whereas, for $k_1>0$, the Duffing oscillator describes the dynamics of a point 
mass in a double well potential, such as a deflection structure building model.

Considering $x \equiv X/r$ and $\tau \equiv \omega_1 t$ 
($\omega_1 \equiv \sqrt{\frac{k_1}{m+M}}$), the equations of motion are 
rewritten in the following form:
\begin{eqnarray}
{\ddot x} + \beta_1 {\dot x} - x + \gamma x^3 & = & \varepsilon
\left({\dot\varphi}^2 \sin\varphi - {\ddot\varphi} \cos\varphi \right), \\
{\ddot \varphi} + \beta_2 {\dot\varphi} + \Omega^2 \sin\varphi & = & \alpha 
- {\ddot x} \cos\varphi. 
\end{eqnarray}
for $\beta_1 \equiv \frac{c_1}{(m+M)\omega_1}$, 
$\gamma \equiv \frac{k_2}{k_1}r^2$, $\varepsilon \equiv \frac{m}{m+M}$, 
$\beta_2 \equiv \frac{c_2}{mr^2 {\omega_1}^2}$, $\Omega\equiv \frac{\omega_2}
{\omega_1}$ ($\omega_2\equiv\sqrt{g/r}$), and $\alpha\equiv \frac{E}{m r^2 
{\omega_1}^2}$ (source of energy).

These equations of motion correspond to a simplified mathematical model for 
oscillator with a limited power supply. In this case, the source of energy is 
given by a DC motor and the parameter $\alpha$ is associated with its
input voltage.  
 
\section{Arnold tongues and Shrimps}

Many systems exhibit behaviour that can be studied by means of two-parameter 
analysis \cite{linsay89,broer08}. Baesens and collaborators \cite{baesens91} 
studied two-parameter families of torus maps that involve change of 
mode-locking type. They analysed codimension-1 and -2 bifurcations for flows on
the torus, and found a large variety of bifurcation diagrams, some of them with
Arnold tongues. Their numerical studies were focussed on resonance regions 
revealing a rich collection of codimension-one and -two bifurcations. 

\begin{figure}[hbt]
\centerline{\includegraphics[height=8.5cm,width=7.5cm]{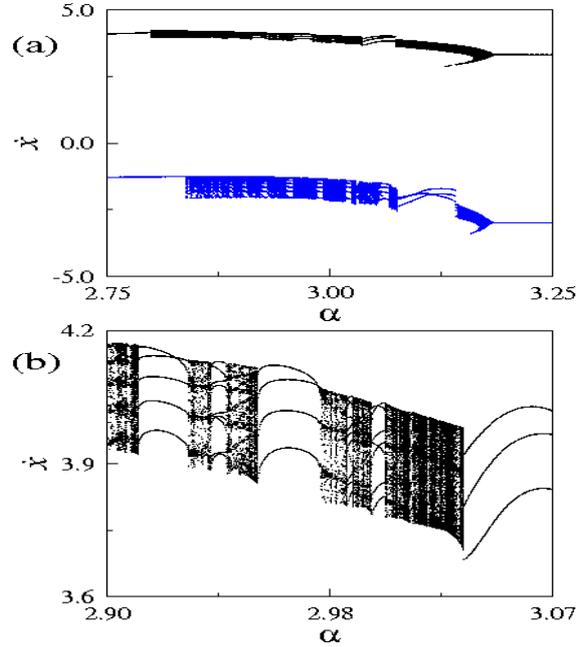}}
\caption{(Color online) (a) Bifurcation diagram showing coexisting attractors 
for $\dot{x}$ in terms of $\alpha$ with $\varepsilon=0.09$. (b) Magnification 
of bifurcation diagram for the attractors plotted in black.}
\label{fig2}
\end{figure}

We analyze, in this Section, self-similar structures and Arnold tongues in the
parameter space of the non-ideal Duffing oscillator were performed by using the
fourth-order Runge-Kutta method with a fixed step. The control parameters were 
fixed at $\beta_1=0.05$, $\beta_2=1.5$, $\gamma=0.1$, and $\Omega=1.0$. We 
consider for dynamic investigations the variations of parameters 
$\varepsilon$ (the ratio of the masses) and $\alpha$ (input voltage of the DC 
motor).  

First of all, we use a bifurcation diagram, as shown in Figures \ref{fig2}(a) 
and (b) for $\varepsilon = 0.09$, to verify possible solutions generated by the
oscillator. This diagram is constructed varying the control parameter $\alpha$.
For each value of the parameter, we plot the  local maximum values of the 
dynamical variable $\dot{x}$ neglecting the transients. As can be seen in 
Figures \ref{fig2}(a) and (b), the bifurcation diagram is composed of periodic 
windows associated with period-adding sequences. See as an example the three 
main periodic windows in (b), with periodic attractor of periods 5, 4, and 3. 
Then, as $\alpha$ is increased the period decreases by 1.

In addition, figure \ref{fig2}(a) exhibits the coexistence of two attractors 
each one plotted with a different color (black and blue). In mechanical 
systems, the coexistence of attractors is quite common non-linear phenomenon 
observed. For example, the coexistence of a large number of periodic attractors
in a mechanical system was observed by Feudel and collaborators 
\cite{feudel98}. They studied the kicked double rotor system, and verified the 
possibility of the system to be stabilized by means of a small perturbation. In
an experimental nonlinear pendulum, it was also observed two coexisting 
attractors \cite{paula2006}. Multiple attractors may be found in many nonlinear
dynamical systems, for instance, driven damped pendulum \cite{gwinn86} and 
spring-pendulum system \cite{alasty06}. 

\begin{figure}[hbt]
\centerline{\includegraphics[height=5.5cm,width=10.0cm]{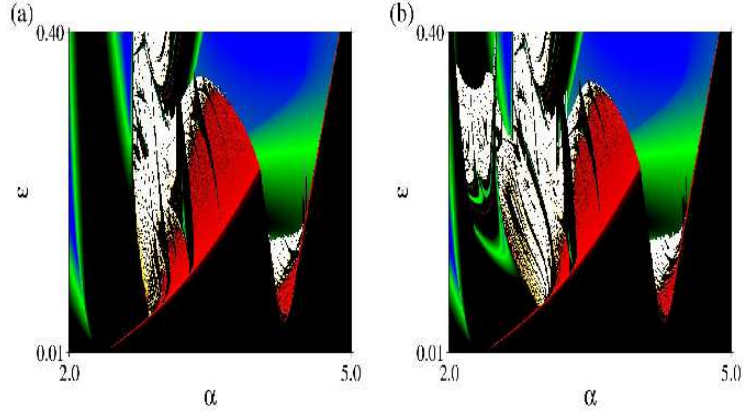}}
\caption{(Color online) Parameter plane diagram for $\varepsilon$ versus 
$\alpha$ with different initial conditions. Periodic solutions are plotted 
in blue, green and black scale ranges, quasi-periodic in red, and chaotic in
yellow and white.}
\label{fig3}
\end{figure}

In order to better characterize the dynamics of the oscillator and to examine 
the structures related to the periodic windows, we construct diagrams of 
two-dimensional parameter space by using the Lyapunov exponents. To evaluate 
these exponents, we use the algorithm proposed by Wolf and coworkers 
\cite{wolf1085}. One positive Lyapunov exponent (LLE) indicates a chaotic 
attractor, all negative exponents a periodic, and two null exponents a 
quasi-periodic or a bifurcation point. 
 
In Figures \ref{fig3}(a) and (b), we plot the parameter plane diagrams, for 
$\varepsilon$ \textit{versus} $\alpha$, using a grid of 800x800 cells. Periodic
solutions are plotted in blue, green and black scale ranges, quasi-periodic in 
red (bifurcation points in red), and chaotic in yellow and white, where the
colors, corresponding to range of Lyapunov exponents values, are introduced
to emphasize the structures details. To make evident the coexistence of 
multiple attractors, these diagrams were constructed considering different 
initial conditions. From an applied perspective, system with coexistence of 
attractor in noisy environments lead to basin hopping 
\cite{desouza-2005,desouza-2007}, the alternate switching among different 
attractors.

\begin{figure}[htb]
\centerline{\includegraphics[height=10cm,width=10cm]{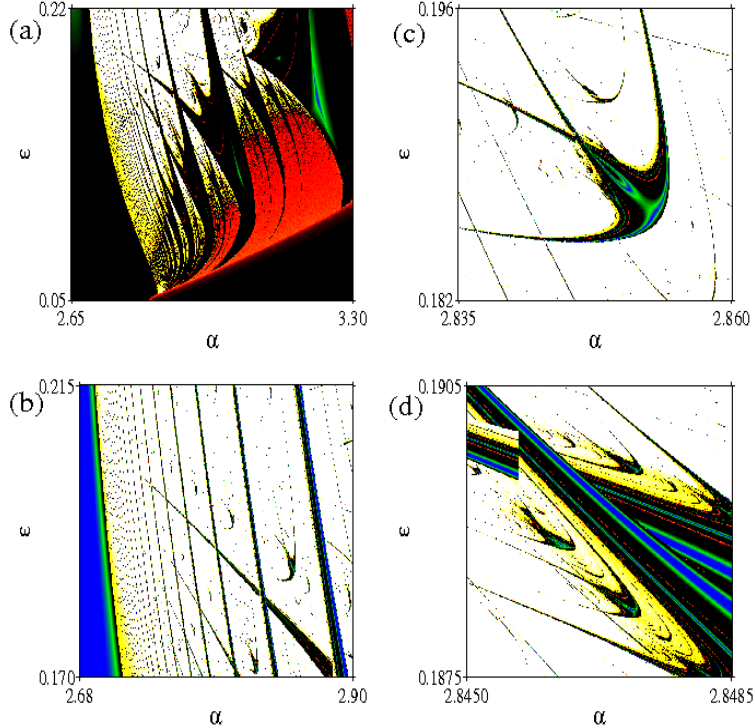}}
\caption{(Color online) (a) Magnification of the parameter plane diagram shown 
in Fig. \ref{fig3}a. (b)-(d) Successive magnifications. Periodic solutions are 
plotted in blue, green and black scale ranges, quasi-periodic in red, and 
chaotic in yellow and white.}
\label{fig4}
\end{figure}

In Figure \ref{fig4}(a), we provide a magnification of rectangular area of 
Figure \ref{fig3}(a) revealing many periodic structures (in black) knowing as
Arnold tongues \cite{desouza-2012} that correspond to phase locking, i.e.,
periodic orbits with the same frequency. Surprisingly, the tongues origins
appear for low value of $\varepsilon$ for a given $\alpha$, accumulate in a 
starting point, namely, the tongues distributions appear highly organized with 
a codimension-2 bifurcation as a point of accumulations. In this case, the 
point of accumulations corresponds to a saddle-node Hopf bifurcation 
\cite{baesens91}.

Moreover, we can observe in Figures \ref{fig4}(b) and (c) small periodic 
structures (blue, green, and black), called shrimps \cite{gallas-1994}, 
embedded in parameter regions with chaotic regimes (yellow and white). The 
shrimps are composed of the central body bordered by a saddle-node and a flip 
bifurcations. As can be noted in Figure \ref{fig4}(d), the shrimp-shaped 
windows present an interesting feature of self-simi\-lar properties. In other 
words, we can verify in successive magnifications (not shown here) the 
repetition of these types of windows for different length scales.

\begin{figure}[htb]
\centerline{\includegraphics[height=5.5cm,width=10cm]{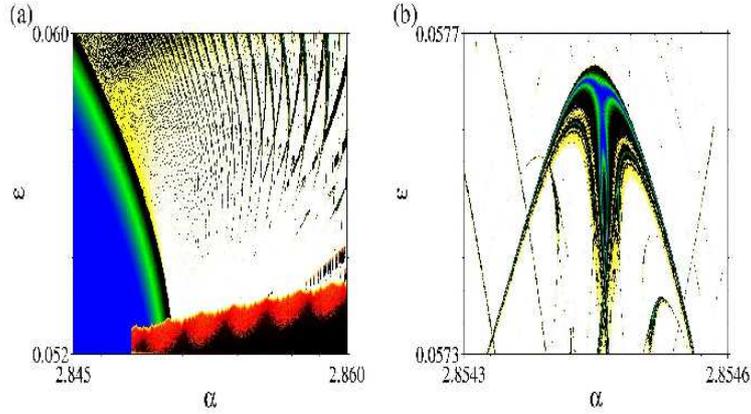}}
\caption{(Color online) (a) Magnification of parameter plane diagram shown in 
Fig. \ref{fig3}a. (b) Magnification of (a).}
\label{fig5}
\end{figure}

In the end, we provide in Figure \ref{fig5}(a) a magnification for a 
rectangular area involving the accumulation point of the Arnold tongues. The 
inferior part of tongues present unexpected shape being quite similar to the
superior part. In addition, embedded in chaotic regions an uncountable number 
of shrimps as shown in Figure \ref{fig5}(b). 

In addition, the periodic windows in the parameter spaces can be understood as
areas delimitaded by the codi\-mension-2 saddle-node Hopf bifurcation curves. 
On the other hand, the Arnold tongues appear whenever the ratio between the 
driven and natural frequencies is a rational number, and they accumulate in a 
codimension-2 point.

\section{Conclusions}

We have investigated the parameter space of a non-ideal Duffing oscillator, 
namely, the mass-spring-pendulum system. Initially, we have identified 
coexisting attractors with period-adding cascades. In this case, it is possible
to verify periodicities of the windows by using the Fibonacci rule 
\cite{desouza-2012}. This rule characterizes an integer sequence, in that the 
sequence is the sum of two previous ones. Moreover, this sequence is related
with Golden ration.

When the source of energy was included in the oscillator we were able to 
observe parameter regions identified as Arnold tongues corresponding to mode 
locked and periodic motion with a common frequency. The mode locked occurs when
the combined motion presented in the mass-spring-pendulum driven by a DC motor 
becomes periodic. We have verified that locked and unlocked regions were 
interwoven in parameter space. Our results showed that the tongues origins 
accumulate in a low value of $\varepsilon$ for a given $\alpha$. Furthermore, 
the tongues are organized from a codimension-2 bifurcation as a point of 
accumulations. We also observed shrimp-shaped structures immersed in parameter 
regions with chaotic regimes, and with highly organized distributions. We have 
found shrimps in small range of the parameter space. 

Moreover, it is important to emphasize that characterization of the attractors 
in parameter space of applied systems is useful to choose robust periodic 
orbits and also to evaluate the attractor changes in case of controlling 
chaotic oscillations.

\section*{Acknowledgments}

The authors thank scientific agencies CAPES, CNPq, and FAPESP 
(2011/19269-11). M. S. Baptista also thanks EPSRC (EP/I032606/1).

\end{document}